\documentclass[12pt,preprint]{aastex}
\shorttitle{Interstellar Fluorine}
\shortauthors{Federman et al.}

\received{2004 May 3}
\begin{document}
\title{$FUSE$ Measurements of Interstellar Fluorine\footnotemark[1]}
\author{S.R. Federman\altaffilmark{2},
Yaron Sheffer\altaffilmark{2}, David L. Lambert\altaffilmark{3}, 
and V.V. Smith\altaffilmark{4}}
\footnotetext[1]{Based on observations made with the NASA-CNES-CSA 
{\it Far Ultraviolet Spectroscopic Explorer} ($FUSE$) which is 
operated for NASA by the Johns Hopkins 
University under NASA contract NAS5-32985.}
\altaffiltext{2}{Department of Physics and Astronomy, University of Toledo, 
Toledo, OH 43606; \linebreak 
steven.federman@utoledo.edu; ysheffer@physics.utoledo.edu.}
\altaffiltext{3}{The W.J. McDonald Observatory, 
University of Texas, Austin, TX 78712; dll@astro.as.utexas.edu.}
\altaffiltext{4}{Department of Physics, 
University of Texas at El Paso, El Paso, TX 79968; \linebreak
verne@barium.physics.utep.edu.}

\begin{abstract}
The source of fluorine is not well understood, although core-collapse 
supernovae, Wolf-Rayet stars, and asymptotic giant branch stars have 
been suggested.  A search for evidence of the $\nu$ process 
during Type II supernovae is presented.  Absorption from interstellar 
F~{\small I} is seen in spectra of HD~208440 and HD~209339A 
acquired with the {\it Far Ultraviolet Spectroscopic Explorer}.  
In order to extract the column density for F~{\small I} from the 
line at 954 \AA, absorption from H$_2$ has to be modeled and then 
removed.  Our analysis indicates that for H$_2$ column densities 
less than about $3 \times 10^{20}$ cm$^{-2}$, the amount of 
F~{\small I} can be determined from $\lambda$954.  
For these two sight lines, there is no clear 
indication for enhanced F abundances resulting from the 
$\nu$ process in a region shaped by past supernovae.  
\end{abstract}

\keywords{ISM: abundances --- ISM: atoms --- stars: individual (HD 208440, HD 209339A) --- nuclear reactions, nucleosynthesis, abundances}

\section{Introduction}

While fluorine consists of the single stable isotope $^{19}$F, 
theories of stellar nucleosynthesis have not
yet been able to identify clearly the origin of this element.  Three 
astrophysical sites are currently suggested as possible significant 
sources -- core-collapse supernovae (SN~II), Wolf-Rayet 
stars, and asymptotic giant branch (AGB) stars.  
In SN~II, the enormous flux of neutrinos 
escaping the core interact with $^{20}$Ne in the Ne-shell, causing 
evaporation of a proton or neutron and 
(in both cases) conversion of a small amount of $^{20}$Ne 
to $^{19}$F (Woosley et al. 1990).  This
neutrino nucleosynthesis is often called the ``$\nu$ process''.  
In Wolf-Rayet stars, $^{19}$F is produced inside the 
star mainly at the beginning of the He-burning phase;
this fluorine will be destroyed later in the evolution of the star unless
extensive mass loss removes the star's layers down to the regions where
$^{19}$F has been synthesized and then ejected into space (Meynet \&
Arnould 2000).  For AGB stars, Jorissen, Smith, \& Lambert (1992) 
suggested that $^{19}$F is a by-product of
neutron captures during He-burning; Mowlavi, Jorissen, \& Arnould 
(1998) provided a detailed description of the process.  
In our effort to understand the 
source of fluorine, we present the first measurements of 
interstellar F with the {\it Far Ultraviolet Spectroscopic Explorer} 
($FUSE$) via absorption from F~{\small I} $\lambda$954.

There are very few observations of F outside the solar system.  Observations 
of HF via rotational-vibrational lines in the infrared (Jorissen et al. 1992) 
are limited to the coolest giants.  
More recently, Cunha et al. (2003) measured HF absorption from red giants 
in the Large Magellanic Cloud and $\omega$ Centauri.  They found that 
the F/O abundance ratio declined as the O abundance declined, a trend 
expected for $\nu$-process synthesis of F but 
Wolf-Rayet stars remain a possibility.  
Liu (1998) obtained the only definitive fluorine abundance in a planetary 
nebula (NGC 4361) via [F {\small IV}] $\lambda$4060.  Interstellar F {\small I} 
was detected through the line at 954 \AA\ toward $\delta$ Sco by Snow \& York 
(1981), who used the $Copernicus$ satellite; the measured 
equivalent width of 10.9$^{+4.5}_{-3.1}$ m\AA\ translates into an abundance, 
log~$\epsilon$(F) = log~(F/H) + 12, of $\sim$ 4.0.  The solar system 
abundance is 4.48 $\pm$ 0.06 from meteorites
and 4.60 $\pm$ 0.30 from HF lines in sunspots (Anders \& Grevesse 1989).  
Neufeld et al. (1997) detected 
HF absorption at 121.7 $\mu$m with the $ISO$ satellite 
from interstellar gas in Sgr~B2, a Giant Molecular Cloud 
near the Galactic Center.  Chemical models (see also Zhu et al. 2002) 
indicate that most of the fluorine 
in molecular clouds should be in HF, while analysis of HF excitation suggests 
severe fluorine depletion assuming the solar system abundance is the 
appropriate measure for comparison.  The results on the abundance of 
interstellar F from HF are consistent with those for HCl in this cloud 
(Zmuidzinas et al. 1995), another hydrogen halide.  However, 
the apparent extreme levels of depletion for both species and the unknown 
reference for elemental abundances in the vicinity of the Galactic Center 
make it hard to interpret these results from a cosmochemical perspective.  
Of the suite of observations, only those of Jorissen et al. (1992) provide 
clear evidence for a site of F production: thermally pulsing AGB stars 
that also make $s$-process nuclides.

This paper describes the detection of F~{\small I} absorption toward 
HD~208440 and HD~209339A, which are about 900 pc from the Sun 
(e.g., Pan et al. 2004a).  These stars were chosen for study because 
they belong to the Cepheus OB2 Association, an ideal site to search 
for $\nu$-process products.  The stars in Cep OB2 formed in an 
expanding shell of gas created by previous members of the cluster 
NGC~7160 when their lives ended in SN~II (Patel et al. 1995; 1998).  
Absorption from H$_2$ is also present at 
954 \AA\ and must be removed for an accurate measurement of 
F~{\small I}; our analysis provides a limiting H$_2$ column density for 
new F~{\small I} $\lambda$954 searches.  The focus of our interpretation 
is on abundance ratios because elemental abundances are less 
accurately known, a result of the relatively large uncertainties in 
total hydrogen column densities.  In particular, we examine 
F/Cl, N/O, and F/O ratios.  Column densities for Cl~{\small I} and 
{\small II} and for N~{\small I} come from $FUSE$ spectra, while 
absorption from O~{\small I} and another strong line of Cl~{\small I}
comes from data acquired with STIS on 
the {\it Hubble Space Telescope} ($HST$).  The F/Cl ratio is useful 
in ascertaining the presence of an anomolous F abundance because both 
halogens are expected to have similar levels of depletion 
onto interstellar grains.  The 
ratio of two volatile elements, N and O, is useful for probing the 
overall level of depletion.  With the information gleaned from 
the F/Cl and N/O ratios, we can explore what the F/O ratios say 
about the site of F nucleosynthesis.

\section{Observations}

\subsection{$FUSE$ Measurements}

The stars, HD~208440 and HD~209339A, were observed by 
$FUSE$ for program B030.  As the two brightest among 5 target stars, 
they offered better spectra with higher signal-to-noise (S/N) 
ratios.  They also have lower H$_2$ column density and therefore
higher residual signal at the position of the 
F~{\small I} line at 954 \AA.  The observations on 
HD~208440 took place on 2001, Aug 03 during 3 orbits with a total
exposure time of 9777 s in 15 sub-exposures.  The data for 
HD~209339A were acquired on 2002, Jul 21 during
a single orbit with a total exposure time of 2044 s 
divided into 4 sub-exposures.  Both stars were
observed in HIST mode through the low-resolution aperture.  
The data were processed with CALFUSE V2.4.  
We then rebinned the data by a factor of 4, yielding 
a S/N ratio of $\sim$ 30 at 950 \AA\ and a two-pixel resolution
element of about 0.06 \AA\ with respective resolving powers of
$\sim$ 17,500 and $\sim$ 15,000 for the LiF and SiC segments.

In addition to the F~{\small I} line at 954 \AA, the $FUSE$ spectra 
included other lines of use for our analysis.  The 
weaker F~{\small I} line at 951 \AA, though not detected, 
provided a check on the column density derived from $\lambda$954.  
The $B-X$ 2$-$0, 3$-$0, and 4$-$0 Lyman bands of H$_2$ between  
1042 and 1083 \AA\ were used to infer the H$_2$ column density; 
this column density allowed us to remove the H$_2$ absorption 
in the vicinity of 954 \AA.  The lines Cl~{\small I} 
$\lambda\lambda$1088,1097 and Cl~{\small II} $\lambda$1071 
yielded the total amount of Cl along the line of sight.  
Since the Cl~{\small I} line at 1088 \AA\ is partially blended 
with the $C-X$ 0$-$0 band of CO, we had to determine the CO 
column density as well.  This was accomplished by a simulataneous 
fit to the $C-X$ 0$-$0 and $E-X$ 0$-$0 bands seen in $FUSE$ spectra 
and to $A-X$ bands in STIS spectra.  Finally, 
the (relatively) weak N~{\small I} line at 1160 \AA\ provided 
the nitrogen column density.

\subsection{$HST$ Observations}

We also utilized archival spectra of the two stars from 
$HST$/STIS observations. For HD~208440, the available spectrum
(dataset O5LH0B) was taken through the 
$0.2^{\prime\prime}\times0.2^{\prime\prime}$ slit at $R = 110,000$, 
while the HD~209339A data (from O5C06M) were secured 
with the highest resolution slit
($0.1^{\prime\prime}\times0.025^{\prime\prime}$) 
that provided $R \sim 200,000$.  Pixels, not subpixels, were 
adopted for the spectrum of HD~209339A, thereby giving us a higher 
S/N ratio and a spectral resolution like that for HD~208440.  
These STIS spectra were needed for our analysis
of the weak O~{\small I} line at 1355 \AA, as well as the strong 
Cl~{\small I} line at 1347 \AA\ and the $A-X$ bands of CO.  

\section{ Analysis and Results}

\subsection{Component Structure}

Since the ultraviolet measurements lacked the necessary 
spectral resolution to discern individual velocity components, 
the basis for the component structure was high-resolution 
($R \approx 200,000$) ground-based measurements (Pan et al. 2004a).  
In particular, we adopted the structure from CH $\lambda$4300 
for our synthesis of the H$_2$ bands.  For the atomic species, 
we used the velocity components showing absorption from 
K~{\small I} $\lambda$7699 and the structure seen 
in Ca~{\small II}~K (3933 \AA) for these components.  In particular, 
the $b$-values, which essentially are a measure of the 
turbulent broadening for these species, and relative column 
densities of the Ca~{\small II} lines formed the basis for 
synthesizing the atomic spectra.  The 
Ca~{\small II} structure was adopted for our syntheses of 
all atomic profiles because the species analyzed here are 
likely to be distributed more widely along the line of sight 
than is K~{\small I}.  Like Ca~{\small II}, O~{\small I}, 
N~{\small I}, and F~{\small I} are the dominant ion in neutral 
diffuse clouds while Cl~{\small I} probably represents an 
intermediate situation.  Inclusion of the additional components 
seen in the Ca~{\small II}~K spectrum did not improve the fits.

\subsection{Profile Synthesis}

The features of interest in the $FUSE$ spectra covered most of the 
spectral passband.  For the determination of the H$_2$ column density 
from the Lyman bands, we used three of the four overlapping 
segments LiF-1A, SiC-1A, and SiC-2B.  The spectra from LiF-2B 
were included in the analysis only when the results were not 
significantly different.  The F~{\small I} lines were covered by segments
SiC-2A and SiC-1B, and the N~{\small I} line at 1160 \AA\ 
was present in segments LiF-2A and LiF-1B.  We co-added 
data from different segments to improve the S/N ratio;  
no evidence for degradation of spectra was noticed over the 
short wavelength ranges of interest to us.  For H$_2$, 
uncertainties were inferred from the range of results for 
individual segments.  For the atomic lines, the uncertainties 
were based on uncertainties associated with equivalent widths and were 
determined from line width and rms dispersion about the continuum level.  
Our fits were performed with the code ISMOD, which uses simple
rms-minimizing steps in parameter space to reach a desired level of
acceptable fit, usually down to 10$^{-4}$ in parameter relative values.

Our modeling of H$_2$ was based on a line list maintained by
S. R. McCandliss\footnotemark\footnotetext{
http://www.pha.jhu.edu/~stephan/H2ools/h1h2data/highjsh2vpnvpp0.dat}.
The normalized continuum was determined by a linear fit of the
highest spectral regions for a given segment. The column density of H$_2$
is not especially sensitive to the cloud structure for our sight
lines because it is derived from damping wings
of the $J$ = 0 and $J$ = 1 lines, the two levels containing most of the
H$_2$ column density. We made sure that the H$_2$ model does not go below
the observed flux, i.e., we avoided creating ``emission'' residuals. In
contrast, the stellar flux tends to dip below the fitted H$_2$ profile owing
to the presence of stellar absorption lines that
were not synthesized. These lines, however,
are exponentially extinguished by H$_2$ absorption so that for regions
of very low residual intensity, the fitted H$_2$ profile and the data
tend to agree extremely well $-$ see Figure~1. In spectral regions with
relatively weak attenuation by H$_2$, we eliminated the undulations
from stellar absorption lines by rectifying the continuum over a short
interval up to 1 \AA\ away from interstellar lines of interest with
low-order polynomials.

We used the fact that H$_2$ and CH track each other well 
(Federman 1982; Danks, Federman, \& Lambert 1984) in synthesizing 
the Lyman bands.  There are 4 and 2 CH components toward 
HD~208440 and HD~209339A, respectively.  The relative
velocities and fractions were held fixed in our syntheses, 
but we modified the $b$-values of the components 
to take into account the different masses of H$_2$ and CH.  This
can only be accomplished once the kinetic temperature 
($T_{kin}$) is known; thus $b$-values were fitted along with 
the rotational excitation temperature from the relative populations of 
the $J$ $=$ 1 and 0 levels of H$_2$, $T_{1,0}$.  Such a procedure
is based on the usual assumption that $T_{1,0}$ 
is a fair representation of $T_{kin}$.  The final $b$-value for 
H$_2$ is a composite of the thermal width and the turbulent width, 
the latter produces most of the broadening in CH.  In all, 
the free fit parameters are total column
density, $N$(H$_2$), radial velocity of H$_2$, 
and six rotational excitation temperatures
relative to the ground level, $T_{J^{\prime\prime},0}$.  The results 
of the syntheses for the component structures and for the column 
densities appear in Tables 1 and 2, respectively.  For completeness, 
the $b$-value derived for H$_2$ and turbulent velocity, $v_{turb}$, 
inferred from $b$(H$_2$) and $T_{1,0}$ are also listed in Table 1.  
Since the sound speed for gas at 80 K that is mostly atomic is 
$\approx$ 1.2 km s$^{-1}$, the turbulent motions appear to be 
essentially sonic.  [Further discussion of the results on H$_2$ 
and CO will be presented elsewhere (Pan et al. 2004b, in 
preparation).]

The next step in our analysis was extracting the column density of 
F~{\small I} from the line at 954 \AA.  (The line at 951 \AA, 
which is affected by the presence of H$_2$ absorption to a lesser 
degree, is too weak for detection in many sight lines, including the 
ones studied here.)  Upon application of the H$_2$ results in Tables 
1 and 2 to the H$_2$ lines in the vicinity of 954 \AA, we saw 
excess absorption which we attribute to F~{\small I}.  The model for 
the H$_2$ lines was divided into the spectrum, revealing the 
presence of $\lambda$954 more clearly.  The F~{\small I} absorption 
was fitted to our template for the atomic lines from 
measurements on K~{\small I} and Ca~{\small II}.  Figure 2 
illustrates the steps yielding an estimate of $N$(F~{\small I}) 
for the gas toward HD~208440 and HD~209339A.  
Since both sight lines have $N$(H$_2$) of about $2 \times 10^{20}$ 
cm$^{-2}$, a limiting column density of $3 \times 10^{20}$ cm$^{-2}$, 
when H$_2$ absorption becomes black and F~{\small I} $\lambda$954 
can no longer be detected, can be inferred from our spectra.

It should be emphasized that the H$_2$ model is very robust thanks to its
dominance over almost the entire {\it FUSE\/} spectral range, and
that around the F I line, the continuum is formed by gently varying slopes
from the wings of the $J$ = 0 and $J$ = 1 lines
of H$_2$. Therefore, the actual
curvature of the continuum on both sides of the 954 \AA\ line is also
very robust. In fact, we use the H$_2$ template to rectify the continuum,
instead of applying an arbitrary polynomial within the {\sc iraf/splot}
task. This assures a much better match with the actual data.
The only variables still affecting the placement of the
H$_2$ model are the position on the
wavelength scale and the alignment of the continuum level. We found
that we needed to raise the continuum level of the H$_2$ template by 5\%
in order to avoid generating ``emission'' regions. This fact
reflects the uncertainty involved with the original rectification over the
{\it FUSE\/} segment.  Because we divided
by the H$_2$ profile, large noise spikes
are generated whenever the residual intensity is very near zero, much
larger, in fact, than the noise appropriate to the precise location of the
F I line. The uncertainties in $W_{\lambda}$, therefore, were 
inferred from regions beyond the H$_2$ lines.  
Some excessive noise is apparent next to the F I line in
HD~208440, where the cores of the H$_2$ lines seem to be narrower than
their templates. This could be caused by extra noise in this spectrum,
perhaps because HD~208440 has lower flux levels than those of HD~209339A.
Indeed, the {\it FUSE\/} exposure time was longer toward HD~208440,
but not by a sufficiently large factor to fully compensate for the lower
flux levels.

Despite the noisy conditions after division by H$_2$ line cores, there
is good signal for the 954 \AA\ line in both stars, with $W_{\lambda}$ $\geq$
15 m\AA. The line is present in the individual segments SiC-2A and
SiC-1B prior to their co-addition. Therefore, there is
no doubt that the detection is real, since it is independent of the segment
being used, of pixel position, and of the line of sight being probed.
The only issue that has to be settled is whether the observed lines might be 
stellar in origin, rather than being interstellar. These lines are indeed 
interstellar, because their radial velocity
agrees very well (within 2 km s$^{-1}$) with that
of the H$_2$ lines, and because their width is
as narrow as the other interstellar
lines, while stellar lines are appreciably wider. In fact,
as can be seen in Figure 2, the fit of the interstellar cloud structure
to the observed widths is very good.

As noted above, an important ingredient of our analysis is knowledge 
of the total chlorine column density, $N$(Cl~{\small I}) $+$ 
$N$(Cl~{\small II}).  For Cl~{\small I}, the line at 1097 \AA\ is 
rather weak, while $\lambda$1347 is quite strong.  The line at 
1088 \AA\ is intermediate in strength and therefore apparently very 
useful, but it is blended with a CO band.  By synthesizing the CO 
bands in our $FUSE$ and STIS spectra with the component structure 
from CH, we were able to incorporate 
$\lambda$1088 into our determination for $N$(Cl~{\small I}).  The 
synthesis of the CO bands was done in two parts: fitting the 
observations of the $A-X$ and $E-X$ bands and then including the 
stronger $C-X$ band in the model for CO absorption.  Our 
values for $N$(CO) are $(1.60 \pm 0.06) \times 10^{14}$ and 
$(9.2 \pm 1.0) \times 10^{13}$ cm$^{-2}$ for the clouds toward 
HD~208440 and HD~209339A.  The resulting column densities for the 
chlorine ions and the other atoms are listed in Table 3.  These 
column densities are based on fits with 7 velocity components 
toward both stars.  Only the velocity zero point was allowed to 
vary in the syntheses of atomic profiles; the residuals from 
these fits are indistinguishable from the fluctuations in the 
continuum on either side of the profile.  The atomic spectra appear in 
Figs. 3 and 4.

\section{Discussion}

The column densities in Table 3 were used for the elemental 
abundance ratios displayed in Table 4, which includes ratios for 
the gas toward $\delta$ Sco and for the solar system for 
comparison.  For $\delta$ Sco, values of $W_{\lambda}$ were 
obtained from the literature (Snow \& York 1981 for F~{\small I}; 
Bohlin et al. 1983 for Cl~{\small II} and Cl~{\small I} 
$\lambda$1097; Meyer, Cardelli, \& Sofia 1997 for N~{\small I}; 
Meyer, Jura, \& Cardelli 1998 for O~{\small I}) 
and were converted into column densities 
through a curve-of-growth analysis with a $b$-value of 2.5 
km s$^{-1}$ for the main K~{\small I} component (Welty \& 
Hobbs 2001).  The adopted $b$-value is consistent with the value 
inferred for dominant ions in the main cloud toward $\zeta$ Oph 
(Savage, Cardelli, \& Sofia 1992).  For the solar system ratios, we 
used the results quoted by Anders \& Grevesse (1989) and 
Grevesse \& Sauval (1998) for N, F, and Cl 
and the newly revised solar O abundance (Allende Prieto, Lambert,  
\& Asplund 2001).  Since N~{\small I}, O~{\small I}, and 
F~{\small I} represent the dominant ion in 
relatively dense, neutral diffuse clouds probed by our observations 
and Cl~{\small I} and Cl~{\small II} essentially all the Cl, we do 
not include the ion stage in what follows.

We first consider the F/Cl and N/O ratios, which are measures of 
unusual atomic abundances.  The three 
interstellar values for F/Cl in diffuse clouds are indistinguishable 
from the meteoritic value (0.16) at the 1-$\sigma$ level.  
This is not surprising since both halogens are expected to have 
similar solid-state chemical behaviors.  The interstellar values 
for N/O are also consistent with the ratio of 0.17 found in the 
solar photosphere, as well as the value ($0.18 \pm 0.02$) obtained 
from interstellar surveys on N (Meyer et al. 1997) and O 
(Cartledge, Meyer, \& Lauroesch 2001) with a common 
set of $f$-values.  Moreover, a 
relatively accurate column density for atomic hydrogen (Diplas \& 
Savage 1994) allows us to derive the elemental O abundance, 
$N$(O)/[$N$(H~{\small I})$+$2$N$(H$_2$)], toward HD~208440; the 
result is $(4.2 \pm 0.8) \times 10^{-4}$ compared to the average 
interstellar value of $(3.47 \pm 0.16) \times 10^{-4}$ (Cartledge 
et al. 2001).  A recent independent analysis for this star 
(Cartledge et al. 2004) yielded $(4.4 \pm 1.1) \times 10^{-4}$.  
Since the lines of sight toward HD~208440 and 
HD~209339A have similar O and H$_2$ column densities and similar 
amounts of reddening (0.34 vs. 0.37), we infer that these sight 
lines do not have enhanced depletion of O onto grains.

As seen in Table 4, the interstellar results for F/O are compatible 
with each other, but their weighted mean of 
$(3.4 \pm 1.0) \times 10^{-5}$ is about half the solar system ratio.  
The bottom panels in Fig. 2 illustrate 
this fact by revealing how a column density 
for F~{\small I} based on the solar system abundance for F is not 
consistent with the data for $\lambda$954.  
This suggests that the sight lines have somewhat more F 
(and Cl) depletion compared with the depletion levels 
for O (and N).  The most recent compilation of solar system 
abundances (Lodders 2003) lists meteoritic results for F and Cl 
that are 0.02 dex (5\%) lower than the values adopted here (though 
within the uncertainties for both determinations).  Such a small 
difference does not affect our conclusions.  

When the levels of depletion for Li and K 
(e.g., Knauth et al. 2003), two other 
elements forming relatively volatile compounds, are added to 
our results for N, O, F, and Cl, a trend with 
condensation temperature ($T_{cond}$) emerges.  [The comparison 
is based on the condensation temperatures quoted by Lodders 
(2003).]  Relative to solar system abundances, N and O ($T_{cond}$ 
$\sim$ 150 K) are depleted by a factor of about 30\%, F and Cl 
($T_{cond}$ $\sim$ 800 K) by a factor of 45\%, and Li and K 
($T_{cond}$ $=$ 1000 to 1100 K) by a factor of 85\%.

There is no clear indication of enhanced F abundances 
resulting from the $\nu$ process in SN~II, although the Cep OB2 
Association experienced core-collapse supernovae in the past 
(Patel et al. 1995; 1998).  
Our sample of two sight lines thus may be 
too limited to discern the presence of a localized event.  
Additional F measurements are planned, expanding the sample 
so that about half the association will be probed.  
Observations of interstellar absorption from other elements 
having a $\nu$-process component, such as Li and B (Woosley et al. 
1990), may aid this quest.  The particular enhancements from 
the $\nu$ process involve the isotopes $^7$Li and $^{11}$B.  
The complex component structure seen toward the stars in 
Cep OB2 (Pan et al. 2004a) essentially 
rules out a discriminating search based on the very weak 
Li~{\small I} doublet at 6708 \AA\ and on measurements 
of the $^{11}$B/$^{10}$B ratio.

Data on the elemental abundance of B relative to the 
F abundance appear to offer the best chance to 
seek evidence for the $\nu$ process in interstellar clouds.  
In their model of Galactic chemical evolution in the solar
neighborhood, Timmes, Woosley, \& Weaver (1995)
suggest that $\nu$-process synthesis of $^{11}$B and F may
be the controlling influence on the abundance of these
light nuclides. The computed production rates depend primarily on
the abundances of $^{12}$C for $^{11}$B and $^{20}$Ne for F.
These abundances in the pre-supernova's C- and Ne-shells,
respectively, are reliably predicted. Hence, the B/F ratio
from the $\nu$ process is predictable. The calculations
show that the B/F ratio is increased by a factor of about 30
when $\nu$-process contributions are included,
yielding a ratio like the solar system's for SN~II from
stars with solar metallicity.  Since the $^{11}$B/$^{10}$B 
ratio is about 4 in diffuse interstellar clouds (Federman 
et al. 1996; Lambert et al. 1998), such a large enhancement for 
$^{11}$B from the $\nu$ process should be easily identifiable in 
regions affected by SN~II.

\section{Concluding Remarks}

We presented the first interstellar fluorine measurements 
with $FUSE$ and described the methodology to extract the 
F column density.  The strongest F~{\small I} line 
at 954 \AA\ is blended with lines from H$_2$; a prescription 
for extracting information on F~{\small I} from the blend 
was given.  In the process, we determined 
that the line at 954 \AA\ can be measured as long as 
$N$(H$_2$) was less than about $3 \times 10^{20}$ 
cm$^{-2}$.  The inferred F/O ratio suggests 
that F is more depleted than O, consistent with expectations 
from condensation temperatures.  While our goal to find a clear 
signature for the $\nu$ process from F measurements was not 
achieved in these first observations, we offered suggestions 
for furthering the search.  Additional F~{\small I} 
observations are planned and determinations of the B/F ratio 
offer possibly the best means for seeking evidence for the 
$\nu$ process in interstellar space.

\acknowledgments
S.R.F. thanks the Department of Astronomy at the Univ. of Texas at 
Austin, where much of this paper was written, for its hospitality.  
Observations made with the NASA/ESA {\it Hubble Space Telescope} 
were obtained from the Multiwavelength Archive at the Space Science 
Telescope Institute.  STScI is operated by the Association of 
Universities for Research in Astronomy, Inc. under NASA contract 
NAS5-26555.  This work was supported in part by 
NASA grants NAG5-4957, NAG5-8961, and
NAG5-10305 and grant GO-08693.03-A from the Space Telescope Science 
Institute to the Univ. of Toledo.  V.V.S. was funded through NASA grant 
NAG5-9213 to the University of Texas at El Paso.

\clearpage

\begin{deluxetable}{lccccccc}
%\rotate
\tablecolumns{8}
\tablewidth{0pt}
\tabletypesize{\scriptsize}
\tablecaption{Component Structure for H$_2$ Syntheses}
\startdata
\hline \hline\\
Parameter & \multicolumn{4}{c}{HD 208440} & & 
\multicolumn{2}{c}{HD 209339A} \\ \hline
$V_{LSR}$\ $^a$ (km s$^{-1}$) & $-$11.3 & $-$8.2 & $-$5.3 & $+$1.0 & & 
$-$5.1 & $-$2.3 \\
Fraction $^a$ (\%) &  0.30 & 0.14 & 0.33 & 0.23 & & 
0.39 & 0.61\\
$b$(CH)\ $^a$ (km s$^{-1}$) & 1.3 & 1.0 & 1.4 & 1.3 & & 
0.9 & 1.0 \\
$b$(H$_2$) (km s$^{-1}$) & 1.5 & 1.2 & 1.6 & 1.5 & & 
1.2 & 1.3 \\
$v_{turb}$\ $^b$ (km s$^{-1}$) & 1.3 & 0.9 & 1.4 & 1.3 & & 
0.8 & 1.0 \\
\enddata
\tablenotetext{a}{CH component structure from Pan et al. 2004a.}
\tablenotetext{b}{Obtained from $b$(H$_2$) and $T_{1,0}$ of 76 and 
91 K for HD~208440 and HD~209339A, respectively.}
\end{deluxetable}

\begin{deluxetable}{ccc}
%\rotate
\tablecolumns{3}
\tablewidth{0pt}
\tabletypesize{\scriptsize}
\tablecaption{Results for H$_2$}
\startdata
\hline \hline\\
$J^{\prime\prime}$ & \multicolumn{2}{c}{log $N$(H$_2$)} \\ \cline{2-3}
 & HD 208440 & HD 209339A \\ \hline
0 & 20.04(0.01)\ $^a$ & 19.87(0.01) \\
1 & 20.02(0.01) & 20.01(0.01) \\
2 & 18.41(0.02) & 18.35(0.04) \\
3 & 18.12(0.12) & 18.02(0.07) \\
4 & 16.03(0.18) & 17.41(0.05) \\
5 & 15.09(0.04) & 16.47(0.21) \\
6 & 14.24(0.10) & 14.06(0.06) \\
total & 20.34(0.01) & 20.25(0.01) \\
\enddata
\tablenotetext{a}{The number in parentheses indicates the uncertainty 
in measurement.}
\end{deluxetable}

\begin{deluxetable}{lccccccc}
%\rotate
\tablecolumns{8}
\tablewidth{0pt}
\tabletypesize{\scriptsize}
\tablecaption{Atomic Results}
\startdata
\hline \hline\\
Species & $\lambda$ (\AA) & $f$-value & 
\multicolumn{2}{c}{HD~208440} & & 
\multicolumn{2}{c}{HD~209339A} \\ \cline{4-5} \cline{7-8}
 & & & $W_{\lambda}$ (m\AA) & $N$ (cm$^{-2}$) & & 
$W_{\lambda}$ (m\AA) & $N$ (cm$^{-2}$) \\ \hline
F~{\tiny I} & 954.8 & $8.17 \times 10^{-2}$\ $^a$ & 
$18 \pm 7$ & $(3.3 \pm 1.5) \times 10^{13}$ & & 
$16 \pm 6$ & $(2.9 \pm 1.4) \times 10^{13}$ \\
Cl~{\tiny II} & 1071.0 & $1.50 \times 10^{-2}$\ $^a$ & 
$23.0 \pm 1.4$ & $(1.9 \pm 0.1) \times 10^{14}$ & & 
$17.3 \pm 1.2$ & $(1.4 \pm 0.1) \times 10^{14}$ \\
Cl~{\tiny I} & 1088.0 & $8.10 \times 10^{-2}$\ $^b$ & 
$46.1 \pm 1.4$ & $(9.0 \pm 0.5) \times 10^{13}$\ $^c$ & & 
$38 \pm 3$ & $(8.5 \pm 1.0) \times 10^{13}$\ $^c$ \\
 & 1097.4 & $8.80 \times 10^{-3}$\ $^b$ & 
$7.5 \pm 0.7$ & $(8.5 \pm 0.8) \times 10^{13}$\ $^c$ & & 
$8.8 \pm 0.5$ & $(9.1 \pm 0.6) \times 10^{13}$\ $^c$ \\
 & 1347.2 & $1.53 \times 10^{-1}$\ $^b$ & 
$88.8 \pm 0.8$ & $(14.4 \pm 0.9) \times 10^{13}$\ $^c$ & & 
$74.2 \pm 0.5$ & $(9.8 \pm 0.3) \times 10^{13}$\ $^c$ \\
N~{\tiny I} & 1159.8 & $9.95 \times 10^{-6}$\ $^a$ & 
$16.2 \pm 0.8$ & $(1.55 \pm 0.08) \times 10^{17}$ & & 
$15.0 \pm 0.6$ & $(1.46 \pm 0.06) \times 10^{17}$ \\
O~{\tiny I} & 1355.6 & $1.16 \times 10^{-6}$\ $^a$ & 
$15.1 \pm 0.9$ & $(8.8 \pm 0.5) \times 10^{17}$ & & 
$16.4 \pm 0.4$ & $(9.9 \pm 0.2) \times 10^{17}$ \\
\enddata
\tablenotetext{a}{Oscillator strengths from Morton 2003.}
\tablenotetext{b}{Oscillator strengths for Cl~{\tiny I} from Schectman 
et al. 1993.}
\tablenotetext{c}{An unweighted average yields $N$(Cl~{\tiny I}) of 
$(10.6 \pm 3.3) \times 10^{13}$ and $(9.1 \pm 1.6) \times 10^{13}$ 
cm$^{-2}$ toward HD~208440 and HD~209339A, respectively.}
\end{deluxetable}

\begin{deluxetable}{lcccc}
%\rotate
\tablecolumns{5}
\tablewidth{0pt}
\tabletypesize{\scriptsize}
\tablecaption{Observed Elemental Abundance Ratios}
\startdata
\hline \hline\\
Ratio & HD~208440 & HD~209339A & $\delta$ Sco & 
Solar System \\ \hline
F/Cl & 0.11(0.06)\ $^a$ & 0.12(0.06) & 0.17(0.10) & 0.16(0.03) \\
N/O & 0.18(0.01) & 0.15(0.01) & 0.21(0.02) & 0.17(0.03) \\
F/O ($\times 10^{-5}$) & 3.8(1.7) & 2.9(1.4) & 3.9(2.2) & 6.1(1.2) \\
\enddata
\tablenotetext{a}{The number in parentheses indicates the uncertainty 
in measurement.}
\end{deluxetable}

\clearpage

\begin{figure}
\begin{center}
\includegraphics[scale=0.67, angle=270]{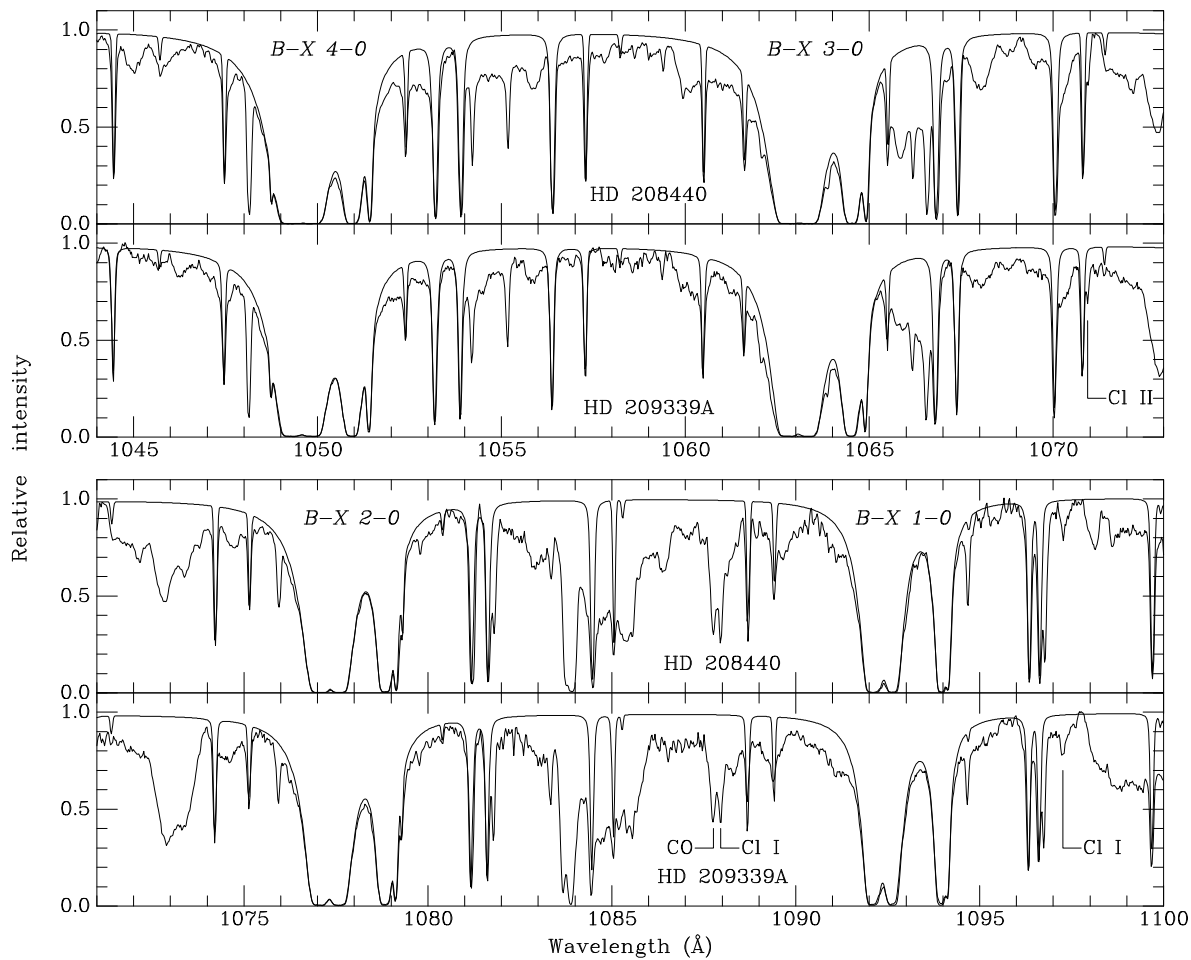}
\end{center}
\vspace{0.3in}
\caption{Global view of four H$_2$ Lyman bands toward HD~208440 
(first and third panels) and HD~209339A (second and fourth panels). 
It is composed of LiF-1A data between 1041--1082 \AA, of
SiC-1A and SiC-2B data between 1082--1087 \AA, and of LiF-2A data
between 1087--1100 \AA. The smooth curve shows the adopted H$_2$ model.
It is based on a simultaneous fit of the 2$-$0, 3$-$0, and 4$-$0 bands for
four {\it FUSE\/} segments (see text). This model yielded the excellent
match for the 1$-$0 band near 1095 \AA.  Stellar features and
other interstellar lines account for the rest of the absorption in the
spectrum.  The Cl~{\small I} and {\small II} lines as well as the 
CO $C-X$ band are indicated in the spectrum for HD~209339A.}
\end{figure}

\clearpage

\begin{figure}
\begin{center}
\includegraphics[scale=0.67, angle=270]{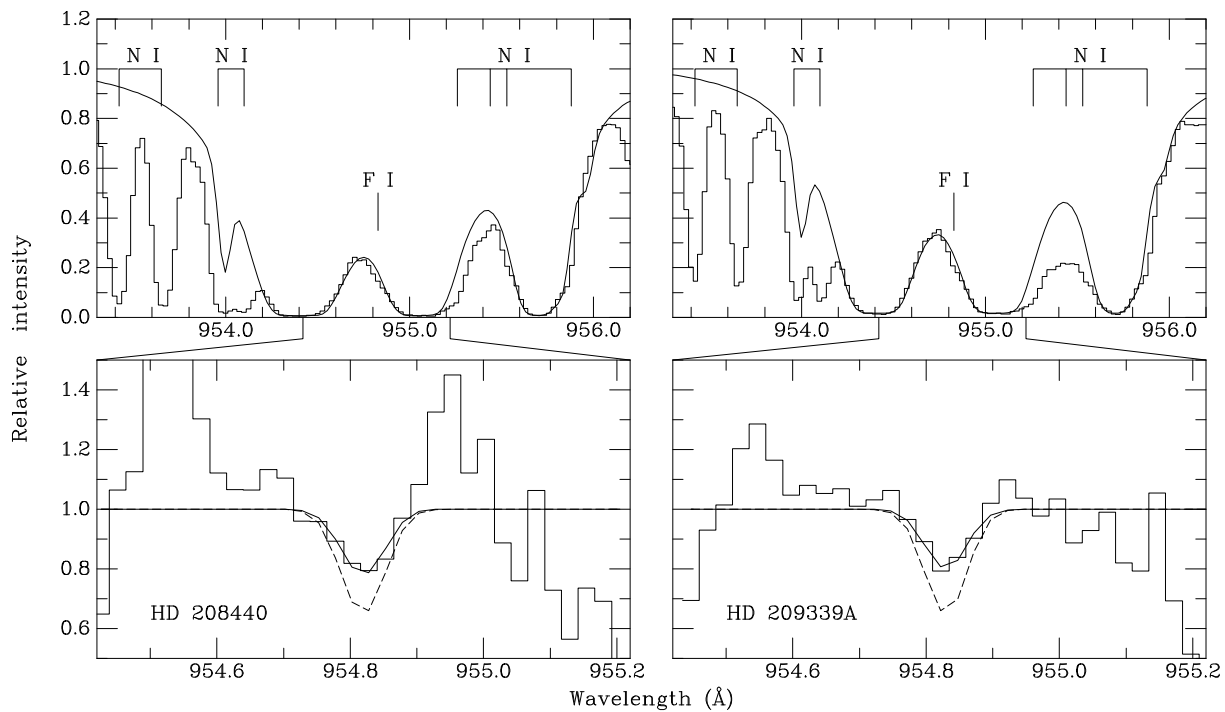}
\end{center}
\vspace{0.3in}
\caption{Spectra showing the steps leading to a determination for
$N$(\ion{F}{1}) toward HD~208440 (left panels) 
and HD~209339A (right panels).
{\it Top panels} -- application of the H$_2$ model from Fig. 1 for 
lines blended with \ion{F}{1} $\lambda$954. The \ion{N}{1} 
lines, which were not fitted because they have substantial 
optical depths at line center but show no damping wings, 
account for the remaining absorption.  {\it Bottom panels} -- an 
expanded view showing the \ion{F}{1} line upon removal of H$_2$ via 
division, resulting in strong residuals at H$_2$ line cores. The
solid curve indicates the absorption associated with the column density
given in Table 3, and the dashed curve the expected absorption for the 
solar system abundance.}
\end{figure}

\clearpage

\begin{figure}
\begin{center}
\includegraphics[scale=0.67, angle=270]{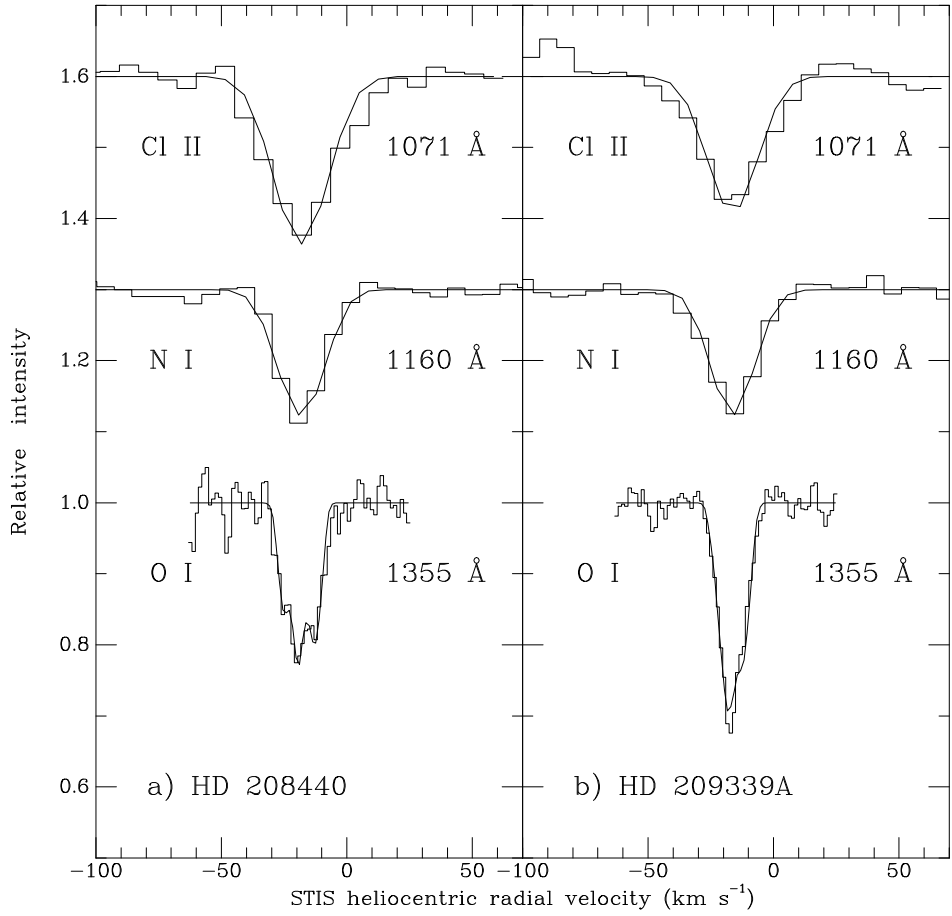}
\end{center}
\vspace{0.3in}
\caption{Interstellar absorption from \ion{Cl}{2} (offset to 1.6), 
\ion{N}{1} (offset to 1.3),
and \ion{O}{1} toward (a) HD~208440 and (b) HD~209339A (histograms) 
with the results from Table 3 presented as smooth curves.
Note the higher resolution of STIS spectra relative to {\it FUSE\/} data. 
The latter were aligned with the radial
velocity of the STIS observation.} 
\end{figure}

\clearpage

\begin{figure}
\begin{center}
\includegraphics[scale=0.67, angle=270]{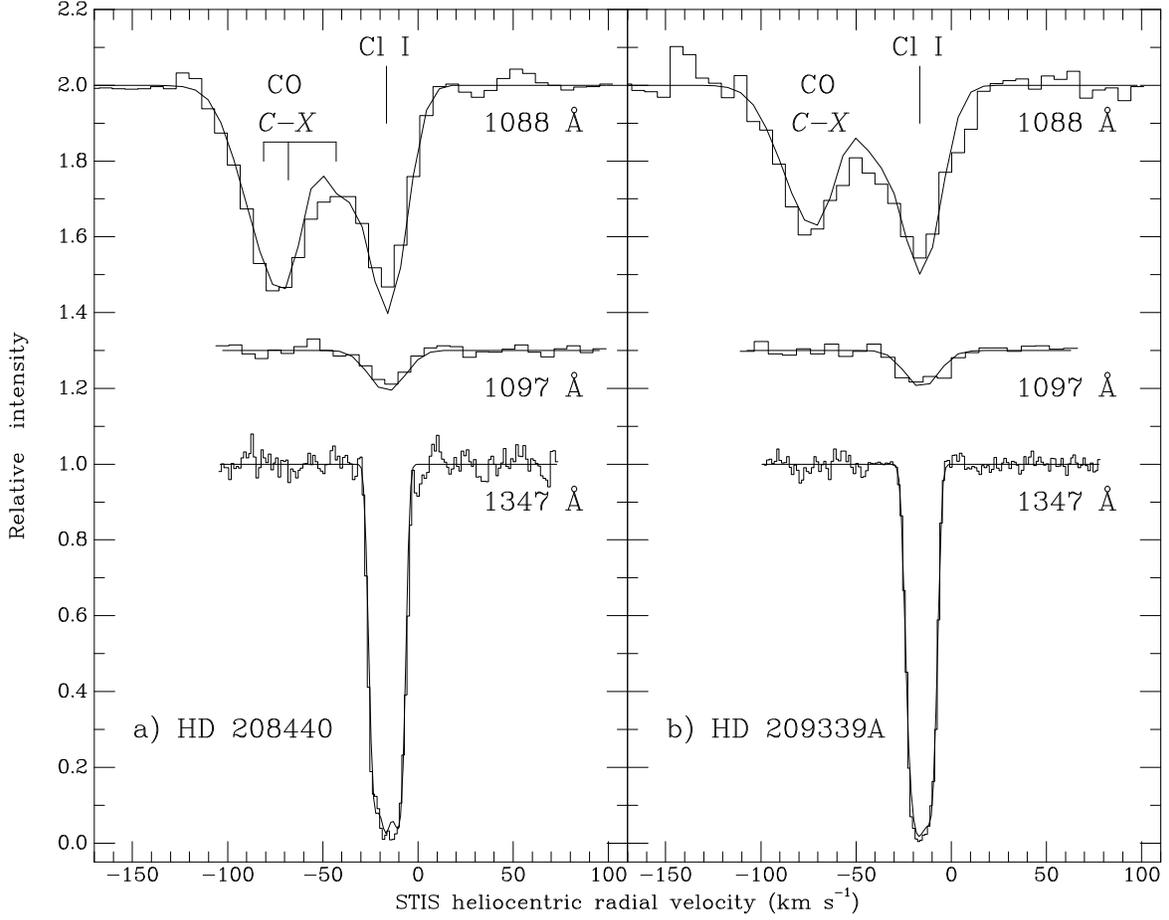}
\end{center}
\vspace{0.3in}
\caption{Interstellar absorption from three lines of \ion{Cl}{1} toward
(a) HD~208440 and (b) HD~209339A.  The spectra for $\lambda$1097 and 
$\lambda$1088 are shifted upward by 0.3 and 1.0, respectively, 
for clarity.  The average results are indicated by
the fits.  The strong $\lambda$1347 line is from STIS
spectra. The $\lambda$1088 line from {\it FUSE\/} is blended with the
$C-X$ 0$-$0 band of CO. The position of the $R$(1), $R$(0), and $P$(1)
lines of CO are shown in the HD~208440 panel. All lines up to $J$ = 5 for
four CO bands, together with \ion{Cl}{1} $\lambda$1088, were 
fitted simultaneously.  For the $C-X$ band toward HD~209339A, only SiC
segments were used.}
\end{figure}

\end{document}